\documentclass[12pt,preprint]{aastex}

\def\etal{{\it et al.}}

\begin{document}

\title{
On the hydrogen emission from the Type Ia supernova 2002ic\footnote{Based on Observations 
Collected at the European Southern Observatory}}

\author{Lifan Wang$^1$, Dietrich Baade$^2$, Peter H\"oflich$^3$, 
J.~Craig~Wheeler$^3$, Koji Kawabata$^4$, Ken'ichi~Nomoto$^5$}

\affil{$^1$Lawrence Berkeley National Laboratory, 1 Cyclotron Rd, Berkeley, CA 94712}

\affil{$^2$European Southern Observatory,
    Karl-Schwarzschild-Strasse 2,
     D-85748 Garching, Germany}

\affil{$^3$Department of Astronomy and McDonald Observatory,
          The University of Texas at Austin,
          Austin,~TX~78712}

\affil{$^4$Opt. \& IR Astron. Div., NAOJ, Mitaka, Tokyo 181-8588, Japan}

\affil{$^5$Dept. of Astrom., University of Tokyo, Bunkyo-ku, Tokyo 113-0033, Japan}

\begin{abstract} The discovery of SN 2002ic 
and subsequent spectroscopic studies have led to the surprising 
finding that SN 2002ic is a Type Ia supernova with strong 
ejecta-circumstellar interaction. Here we show that nearly 1 year 
after the explosion the supernova has become fainter overall, but 
the H$\alpha$ emission has brightened and broadened dramatically 
compared to earlier observations. We have obtained spectropolarimetry 
data which show that the hydrogen-rich matter is highly aspherically 
distributed. These observations suggest that the supernova exploded inside a 
dense, clumpy, disk-like circumstellar environment.
\end{abstract}

\keywords{stars: individual (SN 2002ic) -- stars:
supernovae -- stars: spectroscopy -- stars: polarimetry}

\section{Introduction}
         
SN 2002ic \citep{Hamuy:03, MWW:02, MWW:03} is an important supernova (SN). Spectroscopy
shows that it is a Type Ia SN with strong hydrogen emission arising from
the ejecta-circumstellar interaction. 
Hamuy et al. (2003) point out that SN 2002ic bears a strong spectroscopic
similarity to SN 1997cy, previously identified as a Type IIn \citep{Ger:00, 97cy}.

We present in this {\it Letter} spectropolarimetry observations of SN~2002ic
obtained nearly a year after the explosion. This is the latest epoch for
which spectropolarimetry has been obtained for any SN. We discuss the
nature of the polarization and the structure of the circumstellar environment
surrounding SN 2002ic.

\section{Observations}
We observed SN 2002ic on 2003 July 8.4 (UT), July 29.4 (UT), 
Jul 31.3 (UT), and Sept 29 (UT) using the Very Large Telescope (VLT) of
the European Southern Observatory (ESO) with the FORS1 instrument in 
spectropolarimetry mode. The conditions during the July 8.4  
observations were photometric, and the seeing was around 0.7.$\arcsec$ 
The slit width used in these observation was 1.$\arcsec$ Four separate 
exposures each of 30 minutes were obtained with the retarder waveplate at 
position angles of 0, 45, 22.5, and 67.5 degrees. The July 8.4 (UT) data
provide both reliable spectropolarimetry and spectrophotometry. 
The conditions for July 29.4 (UT) and 31.3 (UT) were less ideal. Only 
exposures at waveplate position angles of 0, 45, and 22.5 degrees 
were obtained on July 29.4 (UT), but all four position angles were 
again observed on July 31.3 (UT). The July 29.4 (UT) and July 31.3 (UT) 
data were combined to produce the spectropolarimetric results. The flux 
calibrations of the July 29.4 (UT), 31.3 (UT) and Sept. 28 (UT) data 
are less reliable but not by more than 10\% as found from the scatter 
of the final reduced data set of each individual exposure. The 
spectral profiles show little evolution during these observations. 
Figure 1 shows the spectroscopic data derived from the July 8.4 (UT) 
observations together with the spectra of SN~1997cy \citep{97cy} 
and SN~1999E \citep{99E} at comparable dates. The most prominent 
spectral line is the broad sharply-peaked H$\alpha$ 
emission that is typical of a Type IIn SN. SN~2002ic remains 
strikingly similar to SN~1997cy and SN~1999E at later epochs \citep{Hamuy:03}. 

\subsection{Time Evolution and Comparisons to Other Type IIn Supernovae}

The H$\alpha$ line in our observations is broader than 
measured in earlier observations \citep{Hamuy:03}. The full width at zero 
intensity (FWZI) can be traced to about 20,000 km s$^{-1}$ compared to
an earlier value of about 4000 km s$^{-1}$. The $H\alpha$ 
line profile shows a strong central peak and is significantly different 
from that of the broad Mg I] 457.1 nm line. Several lines from 
neutral helium and doubly ionized oxygen were also detected. The FWZI of 
these lines are typically around 5,000 km s$^{-1}$. The profiles of 
these lines are actually very similar to the central peak of the 
H$\alpha$ line.  The broad wings, if present, may not be detectable 
because of line blending and the faintness of these lines. 

The flux of the H$\alpha$ line increased by a factor of more than 6 
compared to that of 2002 Nov 11 to 2003 Jan 9. The light curve of 
the H$\alpha$ line 
is shown in Figure 2, together with those of SN~1997cy \citep{97cy} and 
SN~1999E \citep{99E} for
comparison. The H$\alpha$ fluxes derived from our spectra 
are 2.2, 1.9, 
1.8 and 1.7 in units of $\times$10$^{-14}$ ergs cm$^{-2}$ s$^{-1}$ for observations of 
July 8.4 (UT), July 29.4 (UT), July 31.3 (UT), and Sept 28 (UT), respectively.  
 The brightening of 
the H$\alpha$ line after optical maximum is not unique to SN~2002ic, but 
is also observed for similar events such as SN~1988Z \citep{88Z}, 
SN 1997cy \citep{97cy}, and SN 1999E \citep{99E}. The spectroscopic data show
no remarkable evolution during these observations.

\subsection{Spectropolarimetry}

To investigate the details of the polarization 
spectral variation, it is necessary to identify the dominant axis of 
asymmetry in the plot of Stokes parameters, Q versus U, and to decompose the 
data into components along the dominant axis (P$_d$, hereafter) 
and the axis orthogonal to the dominant axis, or the secondary axis 
(P$_o$, hereafter) \citep{Wang:01, Wang:03}. 
The advantage of identifying the dominant axis is that the spectral 
features of P$_d$ will not be sensitive to the uncertainties of interstellar 
polarization (ISP) if the majority of the polarization is produced by a global 
asphericity of the emitting region. A weighted $\chi^2$ fit of the data 
points on the Q-U plot binned to 10 nm shows that the dominant axis is at 
a position angle of $-10^{\rm o}\pm 2.$ 

The polarization due to Galactic dust particles in the field of
SN 2002ic should be less than 0.15\%\ \citep{Heiles02} and cannot be
the cause of the polarization. Dust scattering in the host galaxy 
may contribute. To determine the amount of host interstellar dust 
polarization, we assume that the observed polarization of the central, 
narrow peak of the H$\alpha$ line (from 6553 - 6573 \AA) is due 
completely to interstellar dust. This assumption is justified because
line photons seen in the H-peak are produced by recombination and, thus, are
intrinsically unpolarized (in the absence of large magnetic fields).
In addition, the narrow-line photons, unlike the broad-lines photons,
are not polarized by Thosmson scattering in the circumstellar disk because
they emerge predominantly from regions with negligible
electron-scattering (the thermal velocity of electrons is sufficiently high
that the scattered photons are shifted to the broad wings of the 
H$\alpha$ profile). We thus take the ISP to be that corresponding to
the central 20 \AA of the H$\alpha$ emission peak. This gives an 
estimate of the ISP of about 1.6 percent with
components $Q=0.9\%$ and $U=-1.3\%$. With this ISP 
we determine the net polarization to be about 1 percent for wavelength 
regions outside the narrow H$\alpha$ emission peak.
The dominant and secondary polarization components after correcting for
ISP are shown in Figure 3. It can be seen that across 
the H$\alpha$ line center the dominant component P$_d$ is indeed different 
than that of other wavelength regions. 

The polarization is different from what was found in other SN Ia 
such as SN2001el \citep{Wang:03} which shows a strong feature at 615 nm
corresponding to the strong line of Si II. The polarization in SN 2002ic is
likely to be related to the circumstellar medium (CSM) rather than to the 
explosion process itself.

\section{The Circumstellar Matter Disk}

\subsection{Constraints from H$\alpha$ Emission Line}

At the dates of these observations, SN 2002ic is powered by the high 
energy photons from the ejecta-CSM shock. The velocity of the 
ejecta-CSM shock is found to be around 11,500 km s$^{-1}$ from the 
width of the emission lines associated with the ejecta, such as 
the Mg I] 457.1 nm line and the Ca II IR triplet. As noted above, 
the H$\alpha$ line has a remarkably different profile than the Mg I] line, 
suggesting different origins of these emission lines. Also, 
the He I and the [O III] lines are typically narrower than the
MgI] line with profiles similar to the central peak of the H$\alpha$. 
This is important evidence that the narrow component of the H$\alpha$ and 
the He I lines are produced in the CSM and that the composition of the 
SN ejecta, as reflected in the Mg and Ca lines, is consistent with 
that of a typical Type Ia. 

If the CSM is due to a stellar wind, the luminosity of 
the H$\alpha$ from the unshocked wind is 
$\sim 4.4\times10^{39}{\rm r_{sh,16}}^{-1} {\dot M_{-3}}^2 {\rm u_1^{-2}}$
${\rm erg~s^{-1}}$, where ${\rm r_{sh,16}}$ is the radius of the shock
in units of $10^{16}$ cm and we have taken the emissivity for Case B
recombination at a temperature of $\sim$ 10,000 K. The emissivity is a 
weak function of temperature and changes slightly with
H$\alpha$ opacity \citep{XuMcCray}.
This is a decreasing function of time and cannot explain
the rise of the H$\alpha$ flux. If the CSM were dense enough,
the shocked CSM may become radiative and produce
enough H$\alpha$ emission to account for the observed flux; 
however, the line profile from a thin shell of material moving at 
the shock speed of 11,500 km s$^{-1}$ would be flat-topped and 
hence inconsistent with the observed H$\alpha$ line 
profile. The mass of hydrogen required to produce the observed 
H$\alpha$ luminosity is $\sim$ 6M$_\odot\ \frac{\rm n_e}{10^8 {\rm cm}^{-3}}$.
More realistic modeling of H$\alpha$ emission may alter this estimate
but is unlikely to change the qualitative conclusion
that a few solar masses of dense hydrogen-rich matter 
is required to explain the H$\alpha$ luminosity.

\subsection{Constraints of the Spectropolarimetry}
The configuration we encounter here is different from those for
normal SN~Ia where the polarization is produced by electron scattering
from an aspherical photosphere\citep{Hoeflich}. 
Mass loss from AGB stars is known to be highly asymmetric and 
predominantly on the equatorial plane. The polarization is due to 
scattering by the dense CSM of photons originate from the SN at the 
center. Such systems have been studied in great detail in models of
polarizations of Be stars. In these models the opening angle
of the scattering CSM has the shape of disk with openning angles typically
of the order a few degrees in order to repoduce 
an observed polarization of  order 0.5 - 1\% \citep{McD:01,Mel:01}. Such
a geometry is in good agreement with the required several solar
masses of hydrogen with densities above 10$^{8}$ cm$^{-3}$ that 
contributes to the H$\alpha$ emission. The dense CSM must have a small
volume filling factor.

\section{Ejecta-CSM Interaction and the H$\alpha$ Line Profile}

If the mass loss during the AGB phase were predominantly in the 
equatorial plane, we would expect the shock of the ejecta-CSM 
interaction to be hour-glass shaped. The disk of unshocked CSM 
could be ionized by the hard photons from the ejecta-CSM 
interaction and give rise to the central peak of the H$\alpha$ emission
whereas the shocked CSM could be radiative and give rise to the broad 
H$\alpha$ wings. Detailed calculations are needed to see if this 
scenario can account for the observed line profile and the increase 
of H$\alpha$ luminosity while the overall luminosity decreases.

Another possibility is that broad wings of H$\alpha$ are formed
by scattering, rather than emission from shocked matter.
If the CSM is optically thick to electron scattering, electron
scattering can effectively broaden the H$\alpha$ line and produce 
a center-peaked profile \citep{Chugai:03}.  An example of such a 
profile is shown in
Figure 2. For this simple model spherical symmetry and a temperature
of 30,000 K, presumably produced by the flash ionization of the CSM,
are assumed. Note that this model can produce the smooth line profile
with both the narrow peak and the broad wings whereas such an
agreement would require careful tuning of a model where the
broad wings were shock-accelerated material and the narrow core
were from nearly stationary CSM. In this model, photons
that remain within the core have not been scattered and is thus unpolarized.

It is likely that the flattened CSM is not uniform, but clumpy so 
that the H$\alpha$ line is dominated by emission from the densest clumps.
The clumps may have densities of about $10^{8}\ - \ 10^{9}$ cm$^{-3}$ with 
sizes of about $5\times10^{16}$ cm at which point they would be
marginally optically thick to Thompson scattering and hence provide
the sort of line profiles illustrated by the simple model in Figure 2. 
As noted above, about 1 M$_\odot$ of matter is needed in the clumps/disk to 
reproduce the observed H$\alpha$ luminosity. 

If the disk and dense clumps extend to $\sim$ 200 light days, 
$\sim 3\times10^{17}$ cm, the rising flux of H$\alpha$ could be the result 
of the time delay effect. At early times, only a small fraction of the dense 
clumps would be visible and only at around day 200 when these
observations were made would the majority of the CSM clumps be observed,
therefore giving rise to the increased H$\alpha$ luminosity.
In the mean time, the emission from the ejecta-CSM interaction
should show a constant decrease of luminosity after reaching
maximum a few weeks past explosion, as observed.

\section{Conclusions}
The strong growth in the H$\alpha$ luminosity, the H$\alpha$ line profile,
and the polarization we observe for SN 2002ic are consistent with
several solar masses of clumpy material arrayed to distances of 
$\sim 3\times10^{17}$ cm in a manner that is globally asymmetric. 
The CSM may be related to the dense disks or rings observed in some 
proto-planetary nebulae \citep{Zijlstra, Slijkhuis, Su98}.
The mass loss was found to be highly asymmetric at this stage. 
  For example, IRAS 16342-3814 \citep{Zijlstra} shows a dense torus with mass
0.1 M$_\odot$ and density of $10^8~{\rm cm^{-3}}$. 
Alternatively, a similar configuration may be produced during the 
binary evolution leading to the thermonuclear explosion. During a phase 
of rapid expansion of the donor star very high mass loss may be channeled 
through the outer Langrange points. Wind
interaction \citep{Kwok} can create dense shells or rings that may
naturally explain the observed structure \citep{Kwok,Young:03}.  

The star that exploded as SN 2002ic could be either the 
dying post AGB star \citep{Swartz} or a pre-existing white dwarf 
companion around it.  The large radial extent of the disk seems
to be incompatible with a model in which the hydrogen represents
an expelled common envelope at a distance of $\sim 10^{15}$ cm \citep{Livio:03}.
The proto-planetary nebula phase lasts only for a few hundred years.
This may account for the rarity of events like SN~2002ic, but
raises questions of how the SN is timed to precisely this
phase.  Alternatively, a structure with similar properties may be produced 
during the binary evolution leading to the explosion, although 
questions of the phasing remain.  In any case, events such as SN 2002ic suggest 
that SN Ia can occur in a wide range of CSM environments.

We are grateful to ESO for the generous allocation of observing times. 
We are indebted to Massimo Turatto for providing data of SN 1997cf and 
SN 1999E in digital format.

\begin{figure}
\figurenum{1}
\epsscale{0.8}
\plotone{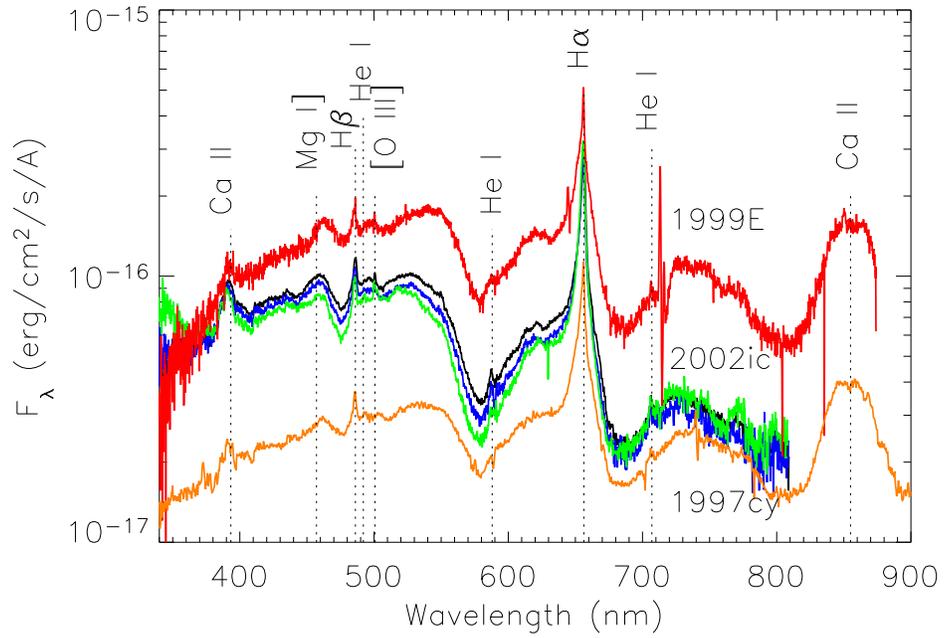}
\caption{The spectra of SN 2002ic are compared with those of SN 1997cy and SN 1999E.
From top to bottom, the spectra are: SN~1999E, SN~2002ic on 2003 July 8, Jul. 31, and Sept 28 (UT),
and SN~1997cy. The SN 1997cy and SN 1999E spectra are 
at epochs where the H$\alpha$ line is close to maximum (see Figure 2). }
\end{figure}

\begin{figure}
\figurenum{2}
\plotone{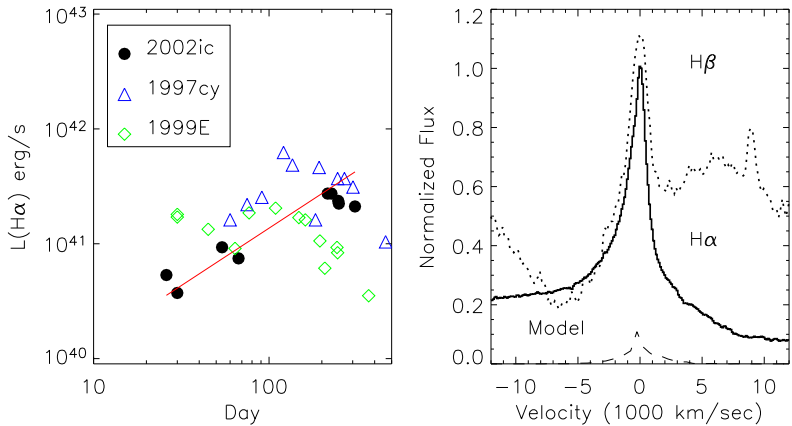}
\caption{(a) The H$\alpha$ luminosity of SN 2002ic compared with those of SN 1997cy 
and SN 1999E. The solid line illustrates a linear increase with time of 
the line flux. SN 2002ic shows comparable H$\alpha$ emission
to SN 1997cy and SN 1999E. The dates of explosion for SN 1997cy and SN 1999E
are uncertain and the data points can shift by more than 20 days on the 
horizontal axis. (b) The H$\alpha$ line profile (solid line) of SN 2002ic is 
compared with the H$\beta$ profile (dotted line). Also shown is a line profile calculated
for a spherical H II region with electron temperature of 30,000 K, and Thomson 
scattering optical depth of 4 (dashed line). The overall agreement of the
line profile suggests electron scattering plays a major role in
forming the line profiles.}
\end{figure}

\begin{figure}
\figurenum{3}
\epsscale{0.5}

\plotone{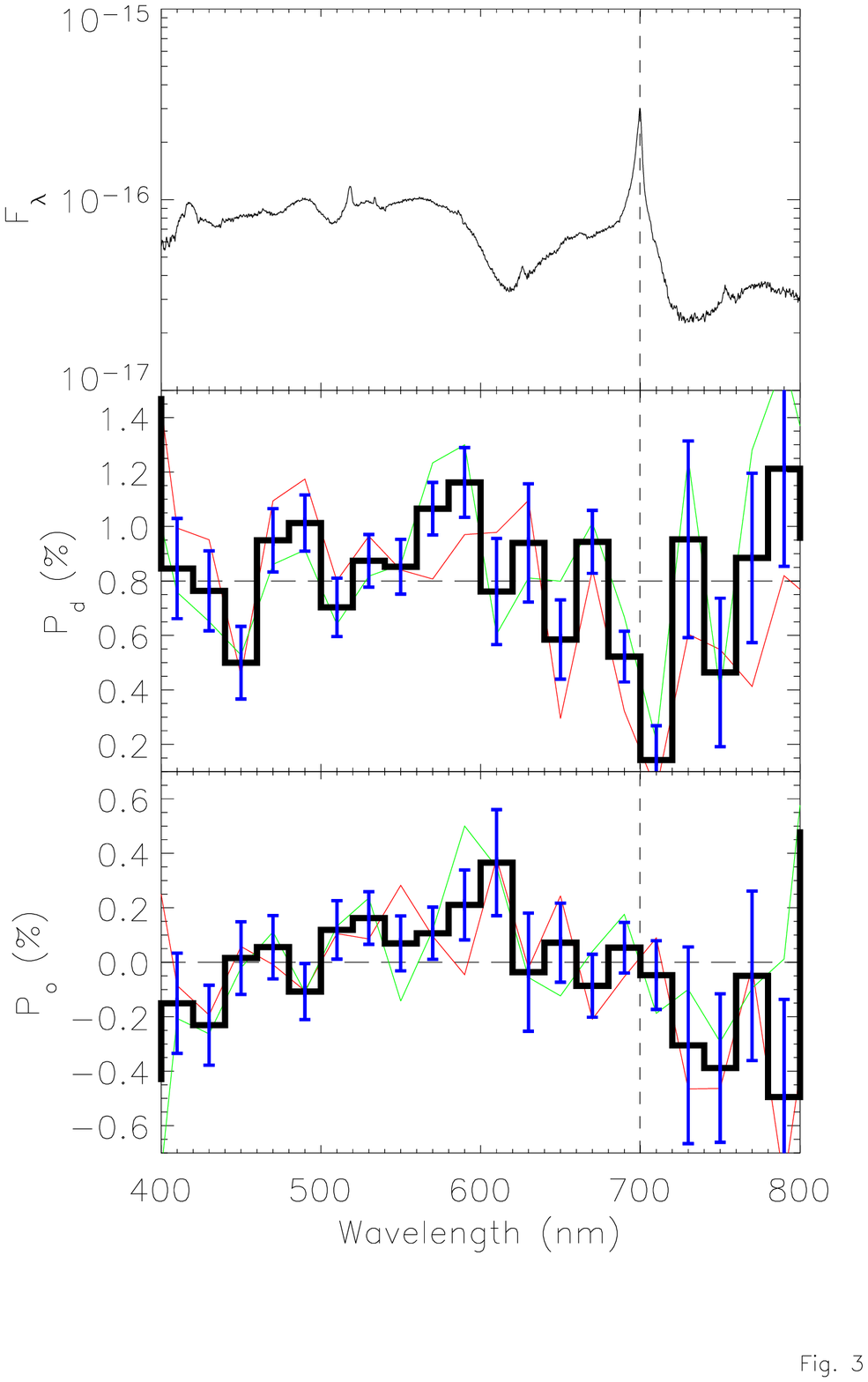}
\caption{The polarization vectors (middle and lower panel) of SN 2002ic 
are shown together with the flux spectrum (top panel). The Stokes 
parameters are projected onto the dominant axis (P$_d$) and the axis 
orthogonal to the dominant axis (P$_o$). The thick solid line shows the 
combined data of all observations, the thin red and green lines show the 
July 8.4 (UT) and July 31.4 (UT) observations, respectively.
The dominant axis is constructed by a weighted linear fit of the data 
points on the Q-U plot, and the position angle of the dominant axis was 
found to be $10{\rm o}\pm0.2$.  In constructing P$_d$ and P$_o$, we have also 
assumed that the central peak of the H$\alpha$ line is 
unpolarized to deduce the ISP component. 
The ISP used in calculating P$_d$ and P$_o$ are 
$Q=0.9\%$ and $U=-1.3\%$. The long dashed lines illustrate
the mean values of the polarization vectors, and the short dashed line 
shows the location of the H$\alpha$ line. The H$\alpha$ line clearly
shows a different degree of polarization when compared to the
overall polarization of the wavelength region from 300 nm to 600 nm.
}
\end{figure}

\pagebreak

\end{document}